# Damage Detection in a laboratory-scale wellbore applying Time Reverse Nonlinear Elastic Wave Spectroscopy (TR NEWS)

E.R. Dauson [a], C.M. Donahue [a], S. DeWolf [b], L. Hua [c], H. Xiao [c], L. Murdoch [b], P.A. Johnson [a]

[a] Geophysics, Los Alamos National Laboratory, Los Alamos 87545, NM, USA
[b] Department of Environmental Engineering and Earth Sciences, Clemson University, Clemson, SC 29634
[c] Department of Electrical and Computer Engineering, Clemson University, Clemson, SC 29634, USA

## 1 Abstract

Time Reverse Nonlinear Elastic Wave Spectroscopy (TR-NEWS) has been used to focus acoustic energy, and make measurements correlated with damage in a variety of industrial materials. Most studies using TR-NEWS in Earth Science have focused on relatively small objects and may have multiple acoustic sources. In Earth, within energy extraction settings, the structure and scale of wellbores makes acoustic focusing challenging. This paper explores the feasibility of applying TR NEWS for damage detection in wellbores by constructing a laboratory-scale wellbore, and using TR to focus and make dynamic linear and nonlinear elastic measurements. After successive cycles of induced, localized mechanical damage the sample, the hysteretic nonlinear elastic parameter $\alpha$, increased with damage cycle indicating progressive mechanical damage. In addition to these strain-dependent changes, TR peak width and changes to peak amplitude near the damage sites was also observed. To deploy TR in a wellbore, it will be necessary to choose sensors that are suitable for the environment, and that can be distributed along the wellbore. Thus, this paper demonstrates that acoustic TR can be conducted using both an intrinsic Fabry Perot interferometer fiber optic strain sensor, and an intrinsic Michaelson interferometer fiber optic strain sensor, as a first step towards deployable sensing for TR in a wellbore.

## 2 Keywords

Nonlinear elasticity, Time Reverse Nonlinear Elastic Wave Spectroscopy (TR-NEWS), NDT, Time Reversal, Fiber Optic Sensors, Wellbores, Acoustics

# 3 Introduction

## 3.1 Motivation

Maintaining and monitoring wellbore integrity is central to many reservoir management strategies, including geothermal, oil and gas, and carbon injection operations. Changes in elastic nonlinearity can be a sensitive measure of mechanical damage in solids—rock, cement, and steel. [1] Nonlinear measurements in metals can provide an indication of pre-damage prior to linear measurements. [2, 3] Similarly, cement and rocks are inherently nonlinear because of their mesoscopic structure. However, inducing additional mechanical damage such as cracks and fractures can increase their nonlinear elastic response significantly, making nonlinear measurements sensitive to damage in all of the materials comprising a wellbore system. [4] From a practical standpoint, measuring acoustic nonlinearity requires sufficient dynamic strain at a point of interest, generally $>=10^{-6}$. [5] Time reversal nonlinear elastic wave spectroscopy (TR NEWS) is one method to focus acoustic energy in a material, to induce locally higher strain, and make local, relative, nonlinear measurements. [6]

This paper explores using TR NEWS in a wellbore to locally measure changes in nonlinear elasticity as an indicator of a change in damage state, by testing this method of damage detection in a laboratory-scaled wellbore. The envisioned deployable system involves statically located acoustic sources in or near the wellbore, and sensing capabilities distributed along the length of the wellbore. By making measurements of relative nonlinearity along the wellbore, autonomously, on a regular schedule, early signs of damage could be detected. This paper includes a demonstration of using time reversal with fiber optic sensors to focus acoustic energy in a laboratory-scale wellbore, as a first step for moving from a laboratory-scale wellbore to a deployable system.

## 3.2 Time Reversal as a method to focus energy

TR is a method to focus energy in complex environments by taking advantage of wave scattering and the reciprocity of acoustic and elastic waves. Briefly, to focus using TR with fixed sensors and sources, a training signal is sent and the response is measured at the sensor. The recorded forward signal gives a measure of how waves propagate between the source and the sensors. This forward signal is reversed in time electronically and reinjected into the source, and because of time invariance in a linear system, sending the reversed signal from the source again will focus the signal at the sensor location. As more sources are added to the system, the focus improves. Alternatively, chaotic cavities or significant medium scattering accomplishes the same thing. Work has been done by many to improve the acoustic focus in space and time at the focus volume, through changes to the time-reversed signal, like using impulse response to increase the amplitudes of late wave components arriving at the focus point, and deconvolution to scale the frequency content of the reversed signal to an approximate delta function made of the frequency content that arrived in the recorded forward signal [7-12].

TR NEWS, the combination of time reversal and amplitude dependent effects, has been used to detect damage both by examining signal amplitude, and looking for strain dependent changes, including harmonic generation and delays in peak time of arrivals [6, 13-23]. For instance, Le Bas et al. used TR to focus energy and detect delamination and a crack on a composite material. By measuring a focused signal inducing strain across a delamination, they observed an increase in strain magnitude in the frequency domain near their input signal center frequency with delamination. Similarly, by measuring a focused signal inducing strain across a crack, they observed an increase in strain magnitude in the frequency domain near their input signal center frequency near the crack. [14] Sutin et al. measured harmonic generation at different strain amplitudes, near and away from cracks in a glass cube. They observed an increase in harmonic generation associated with damage. [17] Payan et al. used TR to focus with different frequency bands and corresponding depths on the surface of heat damaged cement. By measuring the change in time of arrival of the focused peak with strain amplitude, they calculated a local value for the nonclassical, hysteretic nonlinear parameter, α, which is correlated with damage. [22] In most of these examples demonstrating damage detection with TR, the systems were small, or not highly damped. Consequently, there were many scattering wave paths, and highly reflecting surfaces in the system, improving the TR focus.

It is well established that, thanks to strong wave scattering applying even with a single source, it is possible to obtain a high strain focus at a specific point, with low strain elsewhere in the system. [24] This is improved by using multiple sources, rather than a single source as previously noted, and working in a system with many possible wave paths through reflection. [7, 10] For damage detection measurements in uniform, nearly linear elastic materials, mechanical damage, like cracks, can show up as a nonlinear response under sufficient dynamic strain excitation [4]. Having low strain elsewhere during acoustic focusing, makes it possible to know the location of the source of nonlinearity in an otherwise linear system, because when focusing away from a nonlinear region the strain across the nonlinear region would be sufficiently low enough to not contribute significantly to a strain dependent response at the focus point. This strain localization becomes particularly important to consider when looking for damage in systems with materials that are inherently nonlinear, like sandstone. [25-27] Too much nonlinearity in a system, can, in principle break down the time-reversibility of waves, [28], and reduce the ability to reliably focus an acoustic signal. Additionally, linear and nonlinear damping reduces how far a wave can propagate without being overly attenuated. Damping limits the number of wave paths that lead to a TR focus. TR, however, has been used to make measurement both in nonlinear materials, like sandstone [24] and in highly damped systems [29]. However, these systems were often small, with nearby edges to create reflections.

A wellbore is a large and complex system, with materials that are both nonlinear, and have high damping. Cement has strain dependent nonlinearities, as do rocks, and other natural materials into which a wellbore is drilled.

In this work we explore the ability of time reversal to focus acoustic energy in a system with both the inherent nonlinearities in the system. We do so in an attempt to probe localized mechanical damage and thermal damage from the wellbore, located in the casing cement and surrounding rock.

## 4 Methods

### 4.1 Experiment Set up Overall

We built a laboratory-scale, half wellbore sample to examine TR, while allowing easy access for sensing, and the introduction of mechanical and thermal damage. As shown in Figure 1, the sample is composed of a rectangular block of Berea sandstone. The sandstone block has semi-circular channel carved along its length, off centered. Into the channel was cemented half of a steel pipe (Schedule 40 steel, OD 6.032 cm, ID 5.25 cm) using Class G well cement (GCC, Tijeras plant). The cement cured for more than 120 days before experiments began, to reduce changes in the cement between tests due to curing.

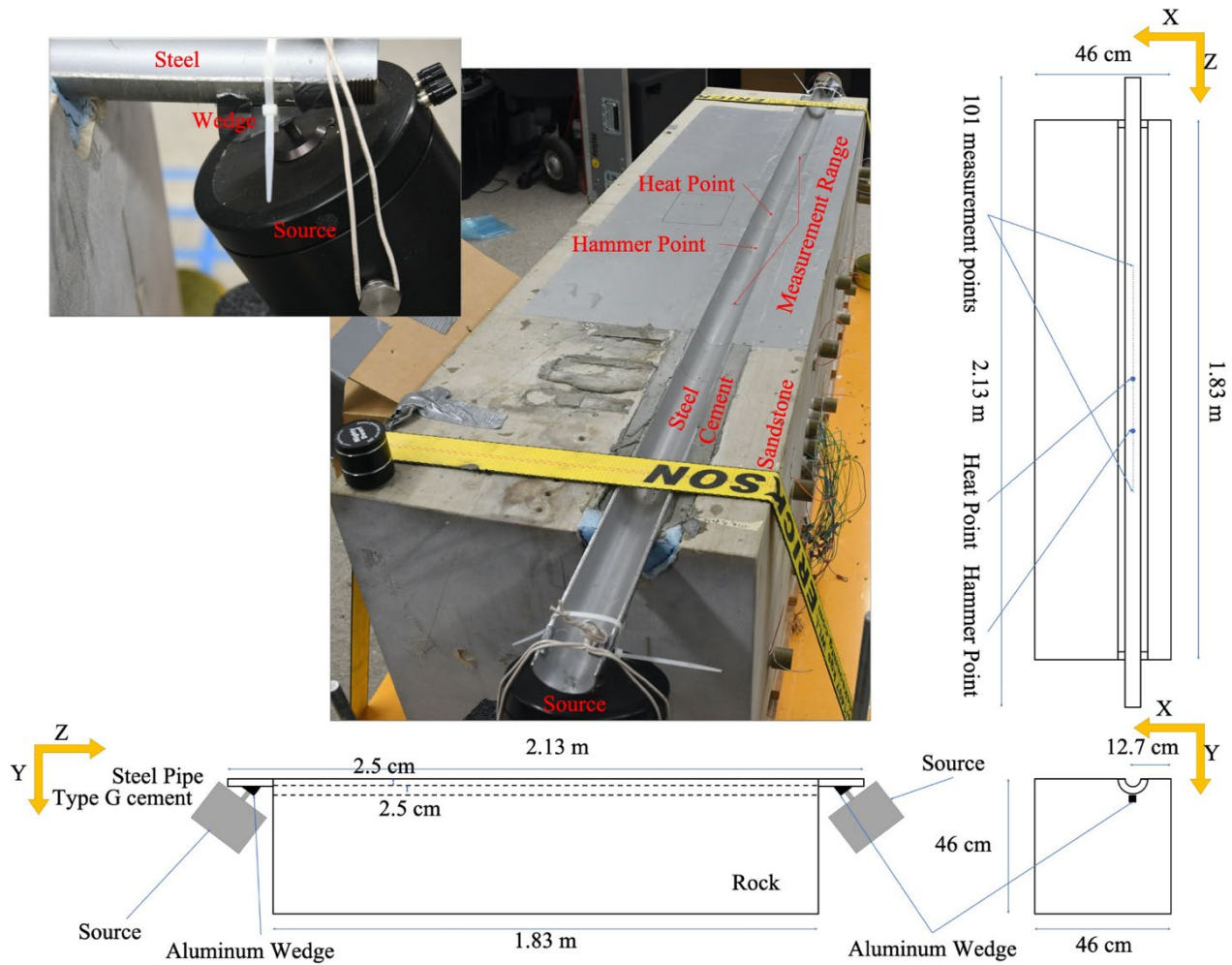

Figure 1: Diagram and photos of laboratory-scale wellbore.

The steel pipe extends beyond the sandstone block on each side by 14 cm and 16.5 cm on the left and right side, respectively. The sources were connected to the underside of the pipe through 5 cm aluminum wedges with a contact surface at a 30° from the bottom of the pipe. Two shakers (Brüel & Kjær Type 4809) were used as acoustic sources. The sources were epoxied (Loctite, Heavy Duty Epoxy) through carriage bolts to the aluminum wedges on either side of the test sample. This wedge-shaped connection between the system and the acoustic sources was chosen to increase reflection, wave paths and wave modes upon entry into the sample. Reflections and alternate paths between the source and the focus point help improve TR focusing [30] by creating additional wave paths between the acoustic sources and the focus point, that would take different lengths of time for a signal to travel. At 87.6 cm and 104.5 cm from the -Z side of the sandstone block, changes were generated though localize heating (heat point) and localized striking with a hammer (hammer point), respectively.

Signals for the acoustic sources were generated using a National Instruments FPGA card 7852R, and amplified with 40dB gain (Brüel & Kjær Power Amplifier Type 2718). The surface velocity in three dimensions was measured at 101 points along the bottom center of the steel pipe, using a 3D laser vibrometer system (Polytec PSV-500, sensitivity 20mm/s), and digitized using National Instrument Digitizer cards 5122.

### 4.2 Time Reversal in the setup

TR was conducted on the laboratory-scale wellbore in two steps. First, a training signal was generated, separately, on each of the two acoustic sources. The response of the training signal, called the forward signal, was measured at one point on the system at a time.

Second, the measured forward signals were time reversed, using one of two methods, described below and simultaneously injected into their respective sources. This second step was repeated at a scaled range of amplitudes. The response of the time reversed, focused signal, could then be measured at the focus point. This was repeated for each of the 101 points along the steel pipe, with TR calculated, then focused, in each of the X, Y, and Z directions.

The training signal used in this experiment was a 380 us linear chirp, from 1 kHz to 20 kHz. The pre-amplification voltage of the input signal was 0.05 V. The training signal and example forward signals measured from both the left and right acoustic sources at the heat point are show in Figure 2. The forward signal was sampled at 1 MHz, and averaged 25 times to reduce noise. The forward signal duration of 16 ms, was chosen to allow the signal to significantly attenuate before the end of sampling, while being limited by the hardware constraints on total sample points per generated signal. Measuring the entire coda improves TR, by providing more information about the transfer function between the acoustic sources and focus point. The signal was produced at a sampling rate of 500 kHz for a total of 8000 sample points. The forward signal attenuates and approaches the system noise level approximately 12 ms after excitation (Figure 2). Assuming a pressure wave speed of 6000 m/s for steel, 12 ms corresponds to approximately 72 m or 40 trips across the sandstone block along the pipe.

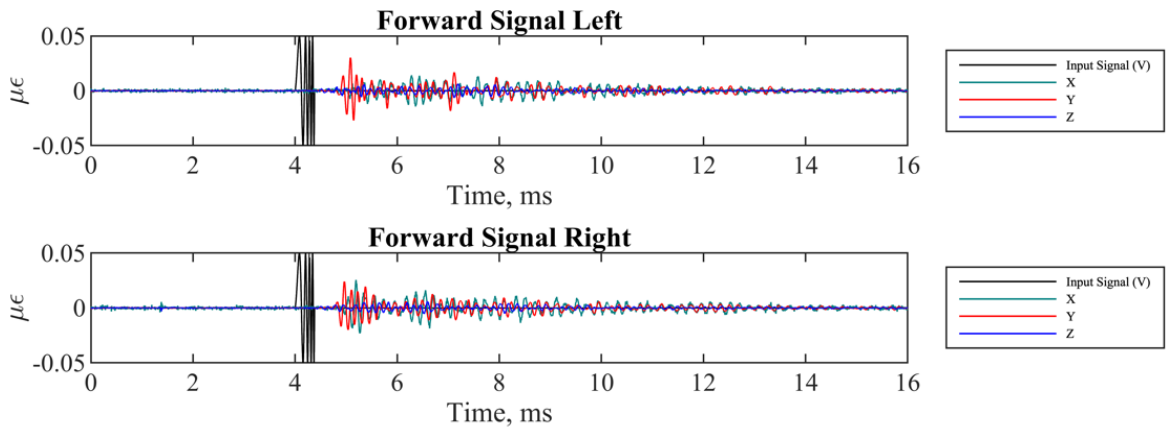

*Figure 2:Input Signal and Forward Response of rock at focus point from left and right sources.*

Two different methods were used for TR in this experiment, namely, deconvolution (D) and impulse response (IR). The methods for IR is described by Anderson et al. [31], as reciprocal TR as "a computed impulse response method." In summary, the input signal was cross correlated with the forward signal in the direction of interest, to obtain an impulse response of the system with reduced noise, and reversed to create a time reversed signal.

The method for D, also known as inverse filtering, begins with the method for impulse response, and then scales the frequency content of the signal, to minimize the effects of system resonances. This method is described by Tanter, [8, 10] and others, as a method to improve the temporal and spatial focus of TR, but with the cost of focus amplitude.[7, 8, 10-12] The "water level parameter," $\gamma$, used in this study was chosen to be 0.9, as optimized for similar applications by Anderson et al. [7]

The time reversed signal for each acoustic source was scaled to 20 pre-amplified voltage amplitudes ranging from -0.25 V to 0.25 V. An example of focus at the range of amplitudes, in the *Y* direction, using deconvolution and impulse response, at the heat point is shown in

Figure 4. The expected focus time, from the calculated signal is 12 ms. As shown in the zoomed-in plot of the focus in *Y* in Figure 4 (middle), the actual time of the peak amplitude of the focus, is strain dependent, suggesting strain dependent nonlinearity in the system. Notice that a focus in the *Y* direction does not correspond to focusing in the *X* or *Z* directions. The negative voltages are a phase inversion relative to their corresponding positive voltages. Consequently, focusing with a -0.25 V signal results in a focused strain in the -*Y* direction while focusing with a 0.25 V signal results in a focused strain in the +*Y* direction.

## 4.3 Localized Damage Test Setup

We created mechanical damage in the laboratory-scale wellbore system, and then used TR to focus energy in each of the three-dimensions at the 101 points in the system to probe these regions. A damage point was created by striking the steel casing with a hammer (Hammer Point, Figure 1). Prior to creating this damage, damage already existed on another point in the wellbore due to localized thermal heating and more information can be found in the supplementary information.

Hammer damage was induced in the sample, by placing a 2.5 cm diameter, 3 cm long aluminum cylinder in the pipe at the "Hammer Point" (Figure 1) and striking the cylinder with a hammer 3 to 4 times. The cylinder aided in directing the impact of the hammer to the bottom of the steel pipe. Data presented below was taken more than five hours after the system was hit for each hammer damage cycle to allow the system time to recover due to slow dynamics [32]. Repeated TR at the focus point immediate after striking, showed that the forward signal changed measurably for 2 hours after the strike time, showing that 5 hours a sufficient time for the properties to not be changing rapidly during measurements.

The hammer striking, TR scan cycle was repeated, similarly to the heating cycles described above for 10 cycles, with no added damage on cycles 7 and 8. The hammer striking visibly damaged the sample, causing cracks to expand near the hammer point. An image of the hammer point, with reflective tape (used for laser vibrometer), removed to show the cement adjacent to the pipe is shown in Figure 3. The image shows some of the damage induced or expanded, including radial cracks, and cracks parallel to the pipe, and cracks at the pipe-cement interface. Cracks appear both parallel and perpendicular to the pipe. There were already small cracks perpendicular to the pipe, throughout the length of the system, appearing during cement curing, but the hammer strikes locally increased the size and number of cracks.

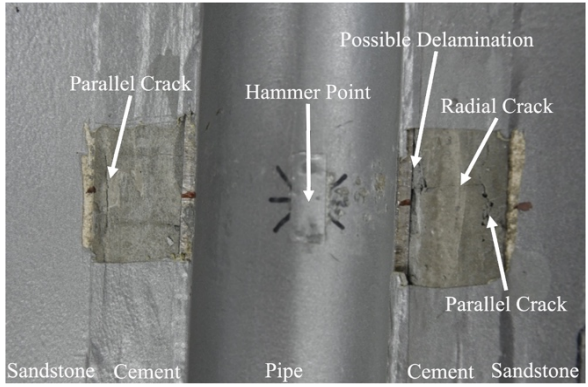

Figure 3: Picture of cracks at the hammer point in the laboratory-scale wellbore system. The radial crack is perpendicular to the z direction.

## 4.4 Fiber Optics Setup

Two different fiber optic systems were used to record the acoustic waves. The first setup, used coherence length gated microwave photonics interferometry (CMPI), as described by Hua et al. [33, 34]. It included a portion of single-mode fiber with two fast laser inscribed weak reflectors, separated by 15cm, which formed an intrinsic Fabry Perot interferometer (IFPI), used as a strain sensor. The interrogator for the IFPI was set to allow a 200kHz acoustic detection band, 66-km measurement range, and 10-m spatial resolution. For the IFPI, tests were conduction in a large block (about 4 tons) of Berea sandstone, rather than the laboratory-wellbore. Optical fiber of length 0.5 m containing the IFPI was taped to the outside of the block. One of the acoustic sources described above was connected through aluminum wedges, to the block about 0.5 m away from the sensor. A second source, ETREMA CU18A Ultrasonic Actuator, was connected to the block approximately 0.5 m away from the fiber, and 1 m away from the first source. The training signal for the IFPI fiber test was 16 ms long, and contained a 0.5 ms long linear chirp from 8 kHz to 12 kHz. It was averaged 25 times and sent through a

500 Hz high pass filter to reduce noise, TR was conducted using the deconvolution method, and rebroadcast at a preamplification voltage of 0.05 V.

The second fiber optic sensor was a 25 mm Michelson interferometer (MI), which was a foreshortened version of the long-baseline borehole strain gauge designed by DeWolf et al. [35] It was epoxied to the laboratory-scale wellbore, on top of an approximately 1 cm x 1 cm x 2 cm aluminum block, positioned about 10 cm in the $X$ direction from the heat point. The training signal for the MI fiber test was 8 ms long, and contained a 4 ms long linear chirp from 1 kHz to 50 kHz. It was averaged 20 times to reduce noise, TR was conducted using the deconvolution method, and rebroadcast at 5 preamplification voltages ranging from 0.05 V to 0.25V.

## 5 Time reversal focusing results

### 5.1 Focus in a wellbore

Prior to the introduction of heating and striking damage, TR was used to focus acoustic energy in the laboratory-scale wellbore, in each of the three-dimensions, $X$, $Y$, and $Z$. Examples of strain at the time of focus is shown in Figure 4. Both foci were created using a preamplification training signal of 0.05 V, and a TR signal scaled to 0.05 V. Note that the input signal resulted in a maximum $Y$ strain amplitude of 0.030 με from the left source and 0.025 με from the right source while the time reversed focused signal has a $Y$ strain amplitude of 0.08 με, and 0.21 με for deconvolution (D) and impulse response (IR) respectively. Thus, for the same voltage input magnitude, TR locally increased the strain amplitude by 1.5 (D) to 3.8 (IR) times over the sum of the maximum forward strains.

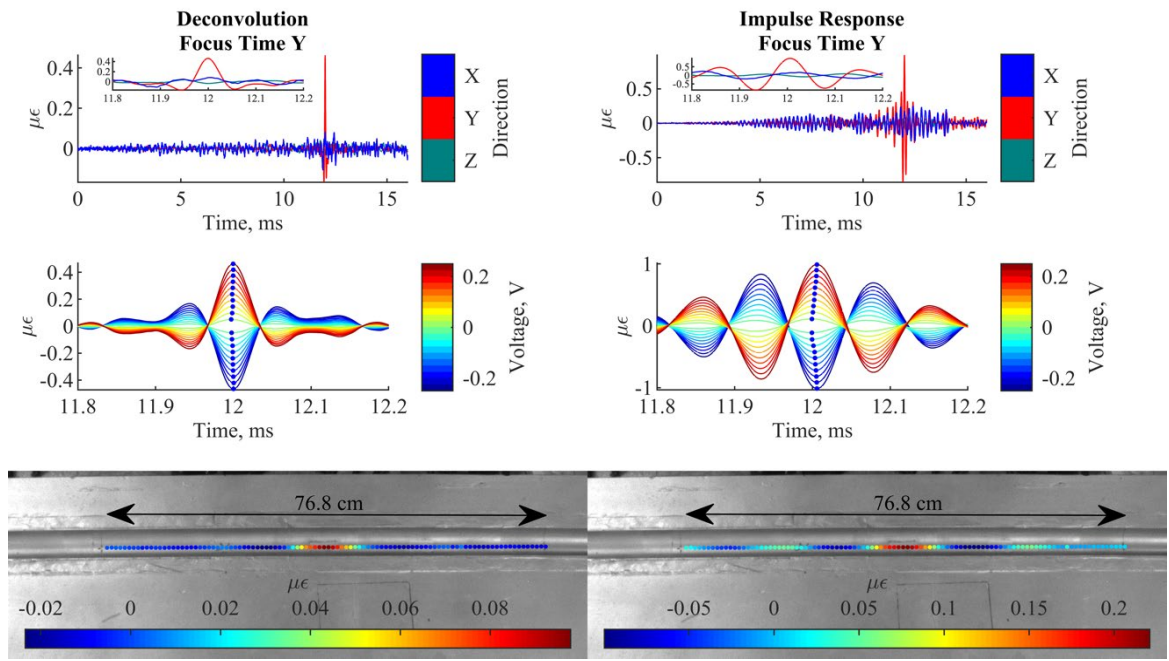

*Figure 4: Dynamic strain in the time domain (top and middle) and strain in the spatial domain (bottom) at time of focus on laboratory-scale wellbore using deconvolution (left) and impulse response (right) focused in the Y-direction. The strain, measured in the X, Y, and Z directions, (top) focused using a 0.25V preamplification voltage. The insets (top insets) zoom in around the focus time. A range of strain amplitudes created by focusing using preamplification voltages ranging -0.25 V to 0.25 V, (middle), shows the change in peak time with strain amplitude.*

For the D-TR method, the focus peak is an impulse with lower amplitude side lobes in both the time domain and the spatial domain, and a peak width (measured by 0 strain crossings) of 65.4 µs and 15.4 cm. For the IR TR method, the strain amplitude is higher, because it does not scale the frequency content to achieve a tight focus, however, in both the time and spatial domains (Figure 4), there are additional side lobes, of higher amplitude.

When focusing in the X and Z directions, time domain signals in general appear similar to those in Y, albeit scaled, to a maximum amplitude of 0.06 µε for D and 0.27 µε for IR, and 0.03 µε for D and 0.13 µε for IR for X and Z, respectively. The time domain pulse widths are also similar. However, in the spatial domain, while the pulse width of the X direction is similar to that of the Y direction, in the Z direction, along the pipe, the pulse width is considerably wider (62 cm). This observation of the pulse width is unsurprising, because much of the energy in the X and Y directions comes from shear waves traveling along the pipe, while much of the energy in the Z direction comes from pressure waves traveling along the pipe. The wave velocity for pressure waves in the system is faster, which results in a longer wavelength, and correspondingly, a wider pulse width.

Localized nonlinear measurements using TR usually make an assumption that strain is sufficiently low to not induce nonlinear elastic effects away from the focus point, so most of the nonlinearity in the focused signal is either from the electronics, and independent of focus location, or from the material nonlinearity itself. [24] Because a wellbore acts as a waveguide, and the system has few reflections away from the pipe there may be some path dependence on the nonlinear response of each point. Figure 5 shows the maximum strain when focusing in the X, Y, and Z directions for the D- and IR-TR methods. This measurement shows the maximum strain amplitudes at all 101 points, over the whole length of the focus signal in time, when focusing at the heat point. Similar to the plot of focus in time (Figure 4) the focus point has the highest strain amplitude. In all directions, for both methods of focusing, the peak strain amplitude at the focus point is about twice the maximum strain amplitude elsewhere.

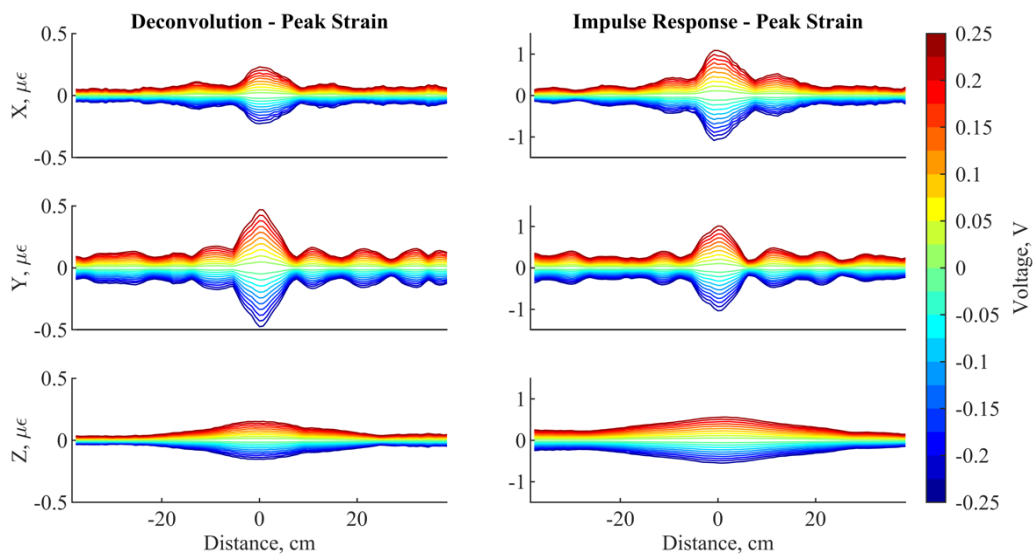

*Figure 5: Peak Strain on route to focus using D (left) and impulse response (right), in X (top), Y (middle), and Z (bottom) directions.*

While all TR scans of the system were tested with both the D and IR TR methods, only the IR TR results are shown below, since the larger amplitude gave a better signal to noise ratio, and more clear results. As seen in Figure 5, despite better localization in time and space at the time of focus (Figure 4) with D, the maximum strain on the route to focus has a similar amplitude relative to peak amplitude with IR.

## 5.2 Focusing using Fiber Optic Sensors

Figure 6 shows normalized strain results from TR focusing using IFPI and MI fiber optic sensors. Both sensors were sufficient to demonstrate TR in the time domain. Additionally, the MI sensor demonstrates a peak time shift with strain amplitude, as would be expected for a measurement on a nonlinear material, demonstrating feasibility for future nonlinear measurements.

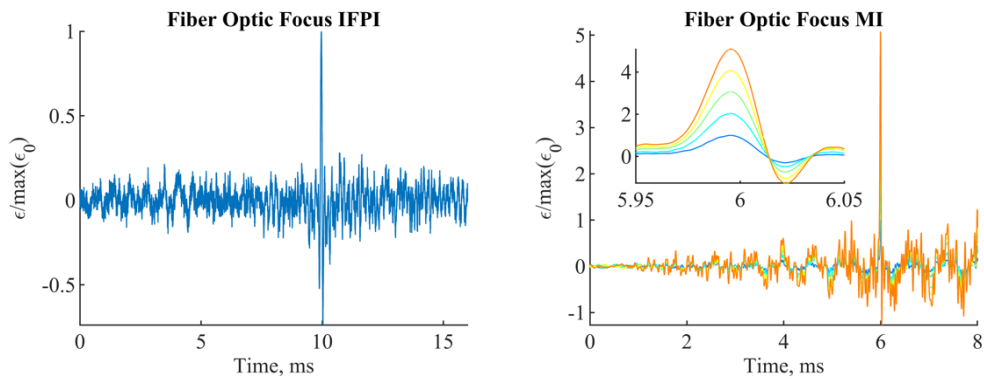

*Figure 6: Time reversal focus using a IFPI sensor (left) and a MI sensor (right), demonstrating the feasibility of using fiber optic sensors for TR in a wellbore environment.*

## 6 Localized Damage Test Results

### 6.1 Strain-Dependent Measurements

Nonlinear elastic response includes changes in wave velocity with increasing strain, which is correlated with damage. [25] Here, to examine the changes in wave velocity, in a similar way to that described by Payan et al. [22], the change in wave velocity ($\Delta c$) is approximated as $\Delta c/c_0 = -\Delta t/t_0$, where $c_0$ is the estimated wave velocity at the 0.05 V TR excitation amplitude, $\Delta t$ is the change in time of arrival in the peak of the time reversed signal, and $t_0$ is the time of flight for a half wavelength path, estimated as the time domain peak width (0 crossing before the peak time to 0 crossing after the peak time) of the 0.05 V TR excitation amplitude signal. Figure 7 shows the changes in wave velocity with strain at a reference point, 35.3 cm away from the hammer point, and at the hammer point, over the striking damage cycles, respectively. If the system has no strain dependent velocity, there would be no change in wave velocity with increasing strain amplitude.

### 6.1.1 Strain-Dependent Wave Velocity

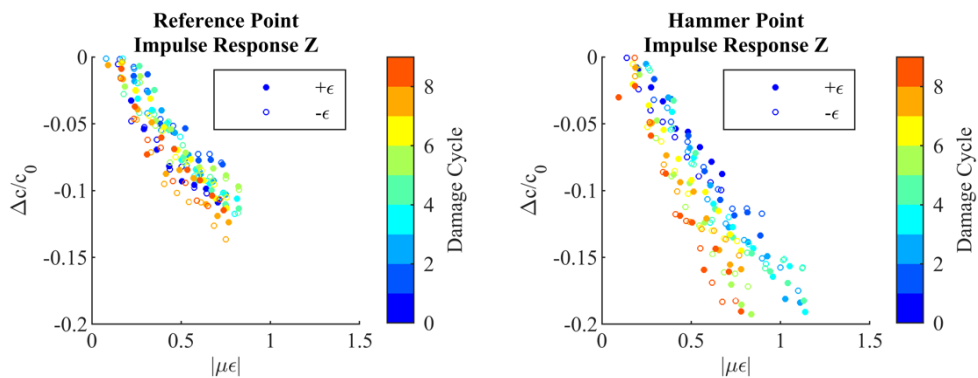

*Figure 7: Peak change in wave velocity and strain amplitude with damage cycle for the impulse response TR method at a reference point 35.3 cm from the hammer point (left) and hammer point (right) in the Z direction. At the hammer point, with increased damage cycles, the amplitude of the slope of change in wave velocity with increased strain increases, indicating an increase in nonlinearity. Changes at the reference point are of smaller magnitude than at the hammer point. Negative strain is induced using TR with a negative input voltage, representing a phase inversion. Differences in the amplitudes of the slope of change in wave velocity with increased strain amplitude in the positive and negative strain indicate differences in behavior under compression or tension.*

Measuring the slope of change in wave velocity verse strain quantifies the hysteretic nonlinearity in the system, with a decrease in slope indicating an increase in nonlinearity. [22] Pristine sandstone and cement both exhibit significant strain dependent changes in wave velocity, [5] so some amount of nonlinearity is measured at all points prior to the induced damage. Additionally, because there may be some amount of path dependence in the laboratory-scale wellbore system, it is not practical to strictly interpret absolute values from a measure of slope of peak time verse strain. However, relative changes indicate changes in the system properties.

The slope of the strain amplitude verses change in wave velocity, can be used as an estimate of the nonlinear parameter α, with $\Delta c/c_0 = \alpha \Delta \varepsilon$ [20]. At the hammer impact point, with increasing damage cycles, nonlinearity increases (Figure 7). This is expected because, in many cases, mechanical damage, like cracks, increases nonlinearity in a system. [4]

### 6.1.2 Hysteretic Nonlinearity, α

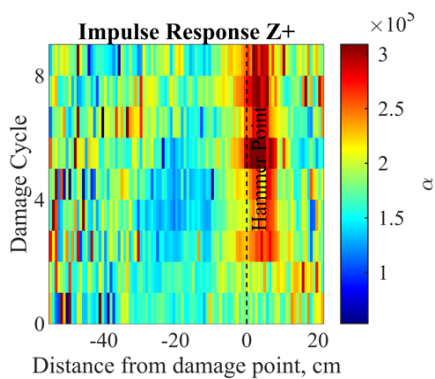

*Figure 8: α with damage cycle in the +Z direction, using IR TR. α increases with damage cycle at the damage point. Plot in -Z direction omitted due to similarity to +Z direction plot.*

Impact cycles resulted in changes to the strain dependent measurements. Figure 9 shows an estimate of α at each of the 101 scan points using IR TR in the Z direction in the system after 9 repeated impact cycles. With repeated hammer strikes, there is an increase in α near the point of damage. An increase in nonlinear elasticity is consistent with the growth of the cracks near the damage as shown in Figure 3. Measurements in the X and Y directions were inconclusive. The Z direction focus should be more sensitive to the radial cracks than the other crack types, because damage detection with time reversal is usually most sensitive to strain across a crack. [19] Minor differences in the +Z and -Z measurements are both a product of measurement noise, and slight differences in the response in these directions. The progressive damage is not seen far from the hammer point, indicating primarily local damage.

### 6.2 Peak Width

In addition to a change of time reversed focus signal amplitude with property changes, other signatures in the time domain signal also change. One, easily measured property of the time domain signal is the time domain peak width (0 crossing before the peak time to 0 crossing after the peak time). Figure 12 shows the peak width for the IR TR method in the +Z direction for hammer impact cycles.

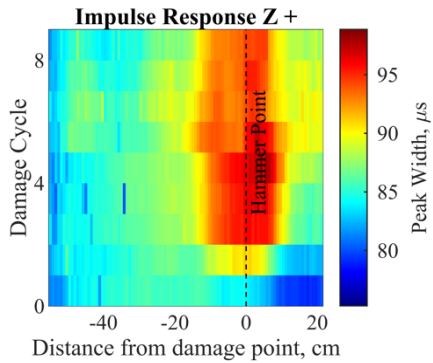

*Figure 9: Width of focus peak (zero strain crossing after focus time - zero strain crossing before focus time) with distance and hammer damage cycle. Near the damage point peak width increases with damage cycle for the impulse response damage method. Plot in -Z direction omitted due to similarity to +Z direction plot.*

At the damage point peak width increases with damage cycle. This gives a spatial band of peak width increase, indicating the local damage. Because peak width is related to the frequency content of the measurement, an increase in peak width is related to a decrease in peak frequency, and related local decrease in wave velocity. While an increase in second harmonic generation (related to the classical nonlinear parameter β) would be expected with damage, the noise in that measurement was too high for unambiguous conclusions.

# 7 Discussion

## 7.1 Localized focusing

Using TR, it is possible to focus acoustic waves in a laboratory-scale wellbore, and there is a significant increase in the focused signal amplitude, when compared to the forward signal amplitude.

The system behaves similarly to a one-dimensional system, in that the pipe acts like a wave guide and much of the energy remains in or near the wellbore itself when the acoustic sources excite from the ends of the steel pipe. In a measure of out-of-plane strain of an input signal traveling from an acoustic source, the measured maximum strain more than 5cm from the pipe is less than 25% of that measured in or within 5 cm of the pipe for the length of the measurement area. While some wave paths exist that focus the energy from reflections off of other surfaces in the system, most are traveling along the pipe, and reflecting in the length of the system. That makes it challenging to get a maximum strain amplitude at the focus point that is considerably greater than twice the amplitude of the maximum strain amplitude elsewhere along the pipe. Any focus strain amplitude above twice the amplitude in the wave path to the focus, requires either different wave types, traveling at different velocities, and, therefore not creating strain at the same time on the pipe path, or wave paths out-of-plane from the pipe. Thus, improvement to the relationship between maximum focus strain amplitude, and strain amplitude elsewhere, may be possible by optimizing different wave modes excitation - e.g., acoustic wedges, shear and compressional transducers - or including acoustic sources that excite waves that do not primarily travel along the pipe, possibly including sources that are not connected directly through the pipe, like cross-wellbore sources, or sources embedded away from the casing.

### 7.2  Localized measurements

In the laboratory-scale wellbore system, it is possible to make localized, relative, nonlinear measurements, which change with property changes. While strain was measured to be relatively high compared to the strain at the focus, the changes in measurements are still localized with the changes in damage and localized relative changes correspond to local property changes, enabling damage detection. However, the ability of TR NEWS to make absolute measurements of nonlinearity in this system has not been determined considering a possible contribution to strain dependence from the wave paths.

### 7.3  Localized property changes

Hammer strikes of the laboratory-scale wellbore system produced progressive local changes in the properties of the system. These property changes resulted in changes in measurements taken using TR to focus acoustic waves. Both strain-dependent and non-strain-dependent measurements changed near these damage points with successive damage cycles.

Consequently, TR can be used to detect changes in properties, and give some indication of an approximate location of the changes. The ability to detect localized changes could be a tool for looking for signs of damage in a full-scale wellbore. This method of TR could enable a user to notice a relative change in properties in a general area, and trigger an inspection. While strain-dependent measurements showed localization, and were sensitive to the property changes, non-strain-dependent measurements, also localized the damage. The measurement of α and the measurement of peak width both are related to local changes in wave velocity, and seemed slightly more discerning for localization than amplitude. Non-strain-dependent measurements take less time, and use less computational memory, because they require only one TR iteration, rather than a range of different amplitudes. It may not be required for a deployed system to make strain-dependent measurements, if non-strain-dependent measurements are sufficient to demonstrate that a change occurred. Peak width, as a measure of frequency peak changes, which is related to nonlinear behavior, seems to balance measurement efficiency with sensitivity.

## 7.4   Damage

In this system, hammering caused an increase in α in the Z-direction. This would be expected for the introduction or expansion of cracks perpendicular to the Z-direction [19]. Some of the damage induced is visible in Figure 3. In addition to cracks perpendicular to the Z-direction, cracks in the cement, parallel to the Z-direction, are also visible. The jaggedness of those cracks could be contributing to the Z-direction measurements, but the cracks also contain a component of the perpendicular the X or Y directions, which might be expected to be measurable with TR in those direction. The hammer point was not prominent with TR in the X and Y directions as changing in localized nonlinearity more than other areas on the system. It is possible that any changes were overshadowed by the permanent damage at the heat point, and would be more prominent when starting from a pristine sample. The measurements were taken at the bottom of the steel pipe in the system, and the hammer was used to induce damage at the measurement site. While cracks on the surface of the cement visibly expanded, the actually damage directly beneath the measurement point could be different than the visible damage.

## 7.5 Time reversal methods

When comparing IR and D-TR methods, D creates a more localized focus, while IR results in greater strain, and more "ringing-up" on the way to focus. Because the wave path in IR crosses any property changes at higher strain, possibly more times, the IR method responds to changes more, so is more sensitive to the existence of changes in the system. For making measurements in the laboratory-scale wellbore, IR TR was better than D TR, but for other applications of using TR to focus in a wellbore, D TR gives a tighter focus in time and space, with fewer side bands, so may be preferable for communication, or creating highly localized perturbations.

## 7.6 Sensing

Primarily, a laser Doppler vibrometer was used for measuring the acoustic response; however, in a full-scale wellbore, utilizing an LDV will not be possible. Fiber optic strain sensors are rapidly being integrated into wellbore for acoustic measurements. Therefore, in this work we show the ability to use TR techniques with fiber optic sensors. Both fiber optic strain sensors can be used to focus acoustic energy. The strain measurement is most sensitive along the fiber, so deployment for TR damage monitoring will need to consider fiber orientation to be sensitive to the damage of interest.

## 8 Conclusions

This paper describes using TR to focus acoustic energy and make acoustic measurements correlated with successive localized damage cycles in a laboratory-scale wellbore. TR is an effective way to locally focus acoustic waves in the system, demonstrating a strain increase of 3.8 times over the sum of the direct forward signal strain amplitudes. Despite most of the acoustic energy remaining near the laboratory-scale wellbore pipe, limiting wave paths, and violating the assumption that strain is low away from the focus point during focusing, it was possible to make localized strain-dependent and non-strain-dependent measurements which picked up the local progression of damage. As a transition towards a deployable system for damage detection in a wellbore, this paper demonstrates TR using fiber optic sensing in the laboratory-scale wellbore environment. A deployable system using TR for damage detection can take advantage of statically located acoustic sources and distributed fiber optic sensing, making a low-cost damage monitoring in a wellbore feasible.

# 9 Acknowledgements

This work was funded by the U.S. Department of Energy Office of Fossil Energy through its Carbon Storage Program as implemented by the National Energy Technology Laboratory. We would also like to acknowledge the contributions of Marcel Remilleux, James Ten Cate, and Luke Beardslee to this work. This work is LA-UR-21-20521.

## 11   Supplemental Information

In the supplemental information we describe the methods and results of heating the sample to create damage.

## 11.1 Localized Heat

Using a (Spot IR Infrared Heaters Model 4150) infrared spot heater, with the head positioned 2.5 cm from the base of the pipe at the "heat point" (Figure 1) the pipe was repeatedly locally heated, and allowed to cool to room temperature, before TR with both the D and IR methods, in each of the X, Y, and Z directions was used to focus on each of the 101 scan points. This was repeated for 14 cycles, with the infrared spot heater at 100% amplitude, for 60 seconds (cycle 1, 2), 120 seconds (cycles 3, 4, 5), 240 seconds, (cycles 6-12, 14), or not heated (cycle 13). The measured temperature at the heating point on the steel side of the pipe reached a maximum of 146.2°C. Between each cycle, the system was allowed to cool to room temperature for 1 hour after heating. To confirm that it had stopped rapidly changing properties, TR was used to examine the heat point every 20 minutes during this cooling period. After an hour, there was no longer a noticeable trend in the difference between the output of the focused signal at the focus point on repeated scans.

## 11.2 Heating Cycles

Localized heating cycles induced progressive localized changes in IR TR measurements. For example, centered at the heat point, the hysteretic nonlinear parameter, α, measured in the Y direction, decreased with thermal cycle. (Figure S1, Figure S2) The peak amplitude of the IR TR signal (Figure S3) and the IR TR peak width (Figure S4) both increased with thermal cycle near the heat point. Nonlinearity and peak amplitude both displayed progressive changes with thermal cycle with an effect extending some distance from the heat point. After waiting 6 months after the conclusion of the thermal cycles, all metrics recovered to some extent, though not completely, towards the pre-thermal cycle values. While these results demonstrate the use of TR to make localized measurements in a wellbore-like environment, it was not possible to guarantee that the thermal cycles actually created changes in the damage state of the system. The recovery indicates that at least some of the thermally induced changes were reversible, and no progressive damage was visible. System changes may have occurred due to localized heat-induced drying, or chemical changes in addition to possible damage. Examining the conditions that might cause a decrease in nonlinearity in the system materials is beyond the scope of this paper.

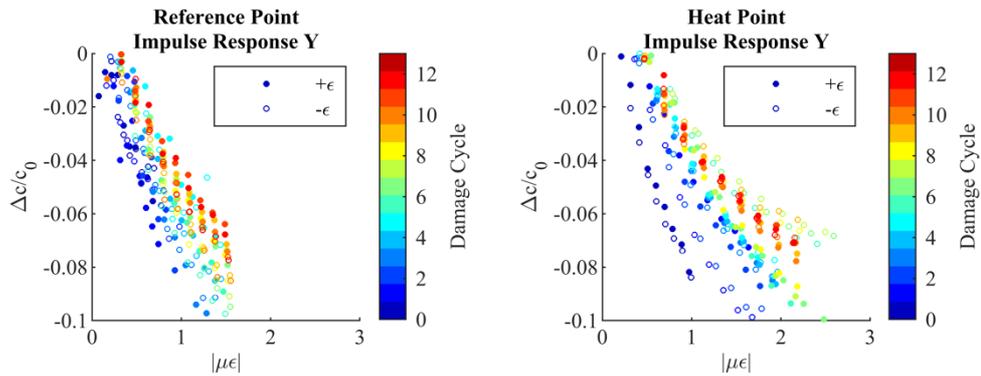

Figure S10: Peak change in wave velocity and strain amplitude with damage cycle for the impulse response TR method at a reference point 38.4 cm from the heat point (left) and at the heat point (right) in the Y direction. At the heat point, with increased heating cycles, the amplitude of the slope of change in wave velocity with increased strain increases, indicating an increase in nonlinearity. Changes at the reference point are of smaller magnitude than at the heat point.

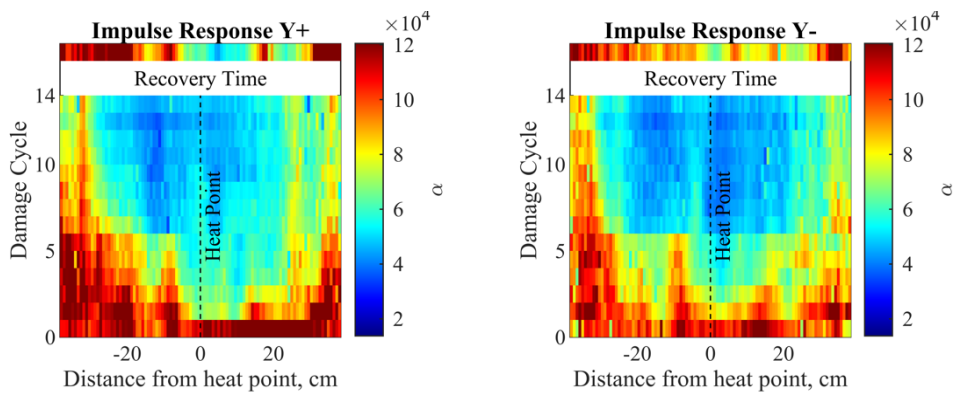

Figure S11: α verse time with damage cycle using impulse response with heating in the +Y (left) and -Y (right) directions. There is a decrease in α with heating cycle centered at the damage point. After an elapsed time of 6 months, α away from the heat point shows some recovery.

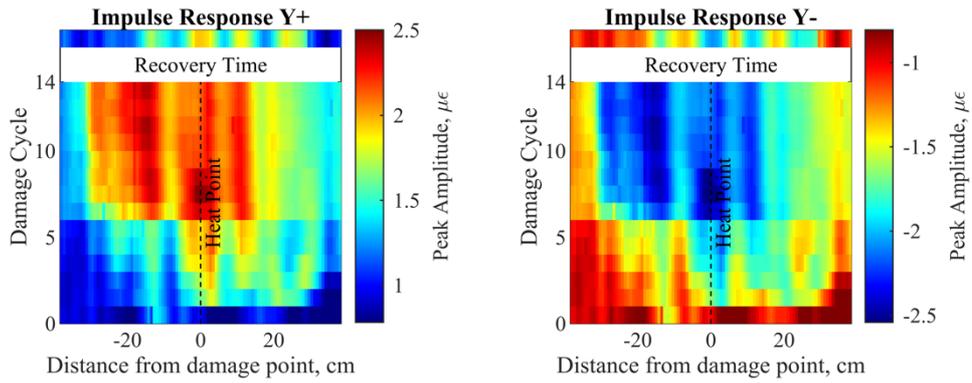

*Figure S12: Peak strain amplitude with damage cycle using impulse response with heating damage. There is an increase in strain amplitude with damage cycle centered at the damage point.*

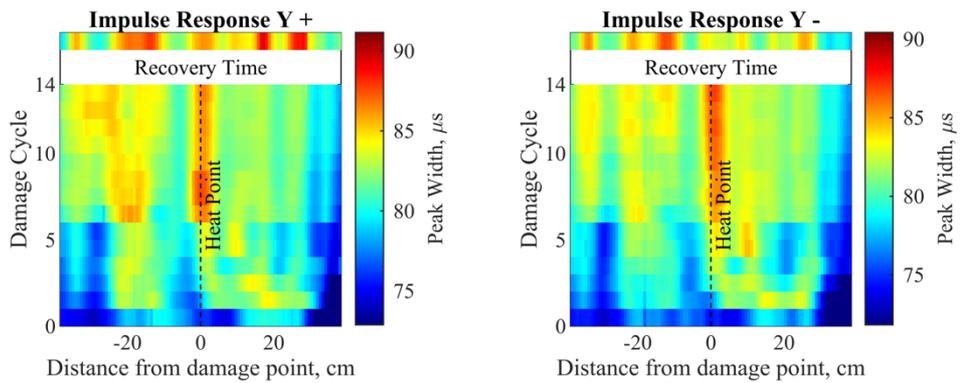

*Figure S13: Width of focus peak (zero strain crossing after focus time - zero strain crossing before focus time) with distance and heating damage cycle. Near the damage point peak width increases with damage cycle.*